\documentclass[12pt,draftclsnofoot,onecolumn]{IEEEtran}
\usepackage{graphicx}
\usepackage{color}
\usepackage{placeins}
\usepackage{float}
\usepackage{tabularx,colortbl}
\usepackage{amssymb}

\usepackage{amsthm}
\usepackage{cite}
\usepackage{amsmath}
\usepackage{caption2}

\usepackage{cases,subeqnarray}
\usepackage{bm,multirow,bigstrut}
\usepackage{textcomp}
\usepackage{latexsym,bm}
\usepackage{booktabs,changebar}
\usepackage{xcolor}
\usepackage{mathtools}
\usepackage{dsfont}
\usepackage{extarrows}
\usepackage{subfigure}

\usepackage{algorithm}
\usepackage{algpseudocode}
\theoremstyle{plain}
\theoremstyle{plain}

\newcommand{\RNum}[1]{\uppercase\expandafter{\romannumeral #1\relax}}
\IEEEoverridecommandlockouts

\begin{document}

%----------------------------title&author&thanks----------------------------
%\title{Trajectory Design for Energy-Harvesting and Hardware Distorted UAV in CF Massive MIMO Systems}
\title{Wireless Energy Transfer in RIS-Aided Cell-Free Massive MIMO Systems: Opportunities and Challenges}

\author{Enyu Shi, Jiayi Zhang,~\IEEEmembership{Senior Member,~IEEE}, Shuaifei Chen, Jiakang Zheng, Yan Zhang,\\Derrick Wing Kwan Ng,~\IEEEmembership{Fellow,~IEEE}, and Bo Ai,~\IEEEmembership{Fellow,~IEEE}

%\thanks{}
\thanks{E. Shi, J. Zhang, S. Chen, J. Zheng, and Y. Zhang are with the School of Electronic and Information Engineering, Beijing Jiaotong University, Beijing 100044, P. R. China. (e-mail: \{jiayizhang\}@bjtu.edu.cn).}
\thanks{B. Ai is with the State Key Laboratory of Rail Traffic Control and Safety, Beijing Jiaotong University, Beijing 100044, China. (e-mail: boai@bjtu.edu.cn).}
\thanks{D. W. K. Ng is with the School of Electrical Engineering and Telecommunications, University of New South Wales, NSW 2052, Australia. (e-mail: w.k.ng@unsw.edu.au).}
}

\maketitle
\vspace{-1cm}
%----------------------------abstract----------------------------
\begin{abstract}
\textcolor{blue}{In future sixth-generation (6G) mobile networks, the Internet-of-Everything (IoE) is expected to provide extremely massive connectivity for small battery-powered devices. Indeed, massive devices with limited energy storage capacity impose persistent energy demand hindering the lifetime of communication networks. As a remedy, wireless energy transfer (WET) is a key technology to address these critical energy supply issues.
On the other hand, cell-free (CF) massive multiple-input multiple-output (MIMO) systems offer an efficient network architecture to realize the roll-out of the IoE. In this article, we first propose the paradigm of reconfigurable intelligent surface (RIS)-aided CF massive MIMO systems for WET, including its potential application scenarios and system architecture. The four-stage transmission procedure is discussed and analyzed to illustrate the practicality of the architecture. Then we put forward and analyze the hardware design of RIS. Particularly, we discuss the three corresponding operating modes and the amalgamation of WET technology and RIS-aided CF massive MIMO. Representative simulation results are given to confirm the superior performance achieved by our proposed schemes. Also, we investigate the optimal location of deploying multiple RISs to achieve the best system performance. Finally, several important research directions of RIS-aided CF massive MIMO systems with WET are presented to inspire further potential investigation.}

\end{abstract}

%\begin{IEEEkeywords}
%Cell-free massive MIMO, unmanned aerial vehicle, wireless power transfer, hardware impairments, harvested energy, spectral efficiency.
%\end{IEEEkeywords}

\IEEEpeerreviewmaketitle

\section{Introduction}

The fifth-generation (5G) wireless network has targeted a $1000$-fold increase in network capacity offering ubiquitous wireless connection for at least $100$ billion devices worldwide, compared with the previous generations of networks.
Recently, with the large-scale commercialization of the fifth-generation (5G) worldwide, the global industry has begun initial research on the next-generation mobile communication technology, i.e., the sixth-generation (6G). One of the key performance indicators for 6G is its extremely massive connectivity for small devices to enable the so-called Internet-of-Everything (IoE) \cite{tataria20216g}.
In practice, most of these IoE devices will be either battery-powered or battery-less due to the associated high-cost of applying conventional power-grid-based solutions.
Unfortunately, the use of limited battery power shortens the lifetime of networks degrading the quality of service. Although frequent battery replacement offers an intermediate solution to this problem, a large number of devices in IoE would further lead to exceedingly high labor and material costs. Therefore, advanced energy replenishing solutions are urgently needed to improve the energy supply challenges of future networks.

Wireless energy transfer (WET) has been proposed to address various practical scenarios where adopting electrical grid is not possible, such as unmanned aerial vehicle (UAV) communications, wireless sensor networks with sensors embedded in challenging environment structures, or inside a human body \cite{hu2020energy}. By exploiting the far-field radiation properties of electromagnetic (EM) waves, the radio frequency (RF) energy signal radiated by the transmitter can be harvested at the receiver which converts it into electrical energy for future use. However, WET technologies face various technical problems such as large path loss attenuation, challenging energy beam alignment, and inefficient resource allocation. Hence, to fully unlock the potential of WET, it must be combined with other advanced communication technologies and architectures to fully unlock the potential of practical IoE networks.

\begin{figure*}[ht]
\label{Figure 1}
\centering
\includegraphics[scale=0.8]{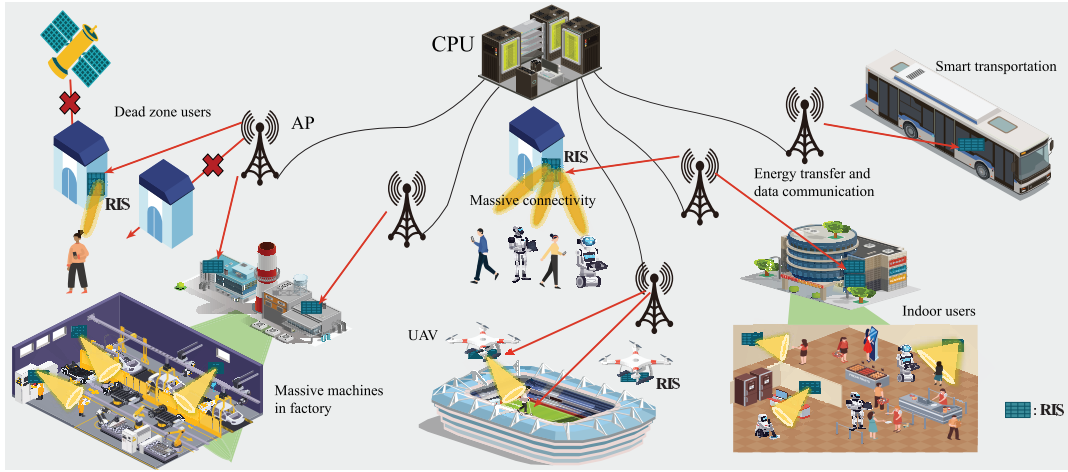}
\caption{Application scenarios of RIS-aided CF massive MIMO systems with WET.}
\end{figure*}

The cellular concept has been introduced over half a century for handling a small number of users in a large area. Recently,  cell-free (CF) massive MIMO systems have been proposed \cite{zhang2020prospective}, which advocates the removal of cellular boundaries for supporting the massive number of users.
In practice, this paradigm can effectively shorten the distance between communication devices and access points (APs) through distributed deployment, thereby improving communication performance.
Specifically, all the APs are connected to a central processing unit (CPU) with high-speed fronthaul links. Different from conventional centralized massive MIMO systems, APs are deployed in a certain range in a distributed manner and cooperate among themselves which offers rich spatial diversity to improve the system spectral efficiency (SE).
On the other hand, the use of high-frequency bands, e.g., terahertz (THz), is expected for 6G networks to cope with the aggressive needs required in massive access \cite{chen2020structured}.
Indeed, by further considering the path loss in high-frequency bands, super-dense APs have to be deployed for reducing communication distances and for ensuring line-of-sight (LoS) between APs and IoE devices.
In general, to support such large-scale multiple access networks, a large amount of energy would be radiated, while the increased interference imposed by super-dense APs has to be carefully managed and controlled. Therefore, it is imperative to study an innovative, spectrally, and energy-efficient, but low-cost 6G wireless network solution.

Recently, reconfigurable intelligent surface (RIS) has been proposed as a promising new technology for reconfiguring the wireless propagation environment through software-controlled signal reflection \cite{tang2020wireless,wu2019towards,di2020smart}. Specifically, RIS requires only low power consumption and low cost. Undoubtedly, RIS can address the shortcomings of CF architecture in future communications and these two technologies complement each other.
\textcolor{blue}{On the other hand, although RIS was initially proposed as a passive component, the coordination of a large number of elements still requires a certain amount of electrical energy \cite{di2020smart}.}
\textcolor{blue}{When there are a large number of RISs, configuring a physical link for each RIS would cause huge resource consumption. Therefore, WET technology is an excellent solution to replace the grid energy-based approach. The energy supply of RIS through WET technology is expected to realize passive deployment of RIS, reduce hardware overhead, and improve the system energy efficiency.}
Despite its great potential, there are relatively little researches on RIS-aided CF massive MIMO systems at present. In fact, some authors have studied the system performance of a single-RIS system, or optimized communication problems in multiple RIS systems \cite{wu2020joint}. \textcolor{blue}{Also, the authors have studied the combination of RIS and WET and the corresponding optimization through advanced optimization \cite{9483903}. The results unveiled the non-trivial trade-off between achieving RIS self-sustainability and the system sum-rate. } \textcolor{blue}{Others have studied the system performance of a single RIS-aided CF massive MIMO systems under idealistic conditions such as with sufficient energy storage and Rayleigh fading channels \cite{van2021reconfigurable}. Besides, the authors introduced a precoding framework for RIS-aided CF networks \cite{zhang2021joint}.} Nevertheless, there is a lack of thorough research on the study of RIS-aided CF massive MIMO systems and their applications with WET technology.

In this article, we try to answer the question: \textsl{How to apply WET technology in RIS-aided CF massive MIMO systems?} To fully exploit the RIS benefits and the WET technology, we design the system architecture applying the WET technology to RIS and discuss its potential future application scenarios. In addition, we design the transmission procedures of the considered system and analyze each procedure. Based on this, we design different hardware architectures of RIS with WET technology for different practical scenarios. Meanwhile, we propose and compare different operation modes of this system, which provide useful insights for the implication of RIS-aided CF massive MIMO systems. With the novel system architecture, we also discuss how to deploy RIS to achieve better system performance. Finally, we highlight potential research directions that deserve further study.

%----------------------------system model----------------------------
\section{Main Application Scenarios and System Architecture}\label{se:model}
In this section, we introduce the main application scenarios of RIS-aided CF massive MIMO systems in future wireless networks and analyze the characteristics of different scenarios. Meanwhile, we provide a detailed introduction to the architecture of the considered system and the corresponding transmission procedure.

\begin{figure}[t]
\label{Figure 2}
\centering
\includegraphics[scale=0.5]{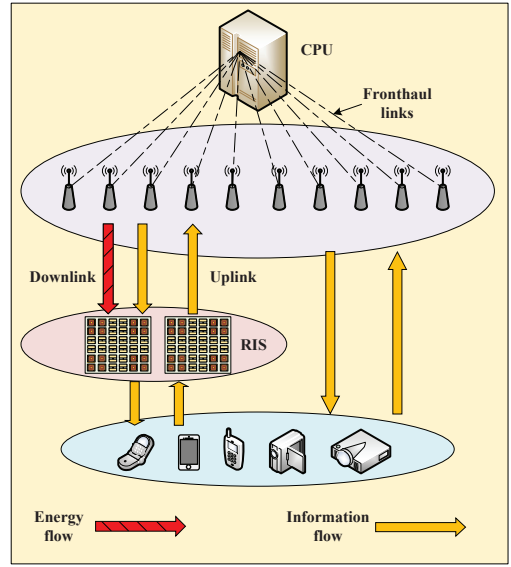}
\caption{System architecture of RIS-aided CF massive MIMO systems with WET.}
\end{figure}

\subsection{Main Application Scenarios}
In the current research, e.g., \cite{wu2019towards,di2020smart} as a passive device, RIS can be flexibly deployed in communication systems for improving the coverage area and providing wireless services for dead spots.
In future 6G network architecture, RIS should be combined with other technologies to achieve better communication performance. Our proposed RIS-aided CF massive MIMO communication architecture serves as an excellent candidate for better exploiting the advantages of WET technology and realizing the vision of IoE.

The future wireless networks are expected to make full use of the low, medium, and high full-spectrum resources to achieve seamless global coverage of space, sky, and earth trinity, such that they can satisfy the stringent demand for establishing unlimited safe and reliable ``human-machine-object" connections anytime and anywhere. Indeed, the success of this desired vision relies on the support of massive access required by the IoE, requiring higher transmission rates, lower delays, and higher reliability \cite{giordani2020toward}.
The main scenes include two categories, densely populated spaces and large-scale factories with densely deployed equipment. As shown in Fig. 1, crowded spaces include large indoor shopping malls, basketball courts, restaurants, stadiums, and so on. In contrast, the scene of a large-scale factory with densely deployed equipment is relatively static and the only main challenge is to facilitate energy harvesting at potential equipment. \textcolor{blue}{Note that in the factory scenario, CPU and AP are internal devices rather than external additional devices.}
The main feature of the former is that the mobility of personnel is relatively high and the wireless equipment in the latter has only limited mobility. For the former scenes, there is generally more uncertainty in wireless communication channels. For example, for the indoor scenario, under the original CF system architecture, we can deploy multiple RISs in some communication dead zones and improve the quality of communication by increasing the number of RISs.
\textcolor{blue}{Since WET technology is exploited to supply the necessary energy to RISs, there is no need to deploy physical power lines for charging which is more flexible in practical implementation. Moreover, in the indoor environment, there is an upsurge in the demand for temporary communication such as large-scale activities in which the role of flexible RIS deployment is particularly prominent.}
As for outdoor environments such as the gymnasium, we can deploy an unmanned aerial vehicle (UAV) on-demand as a carrier of RIS to offer signal coverage and enhancement through intelligent design of the track. On the other hand, for the latter scenes that equipment is not mobile, we only need to deploy the RIS in a fixed location in advance to improve the system performance.

\begin{figure}[t]
\label{Figure 3}
\centering
\includegraphics[scale=0.5]{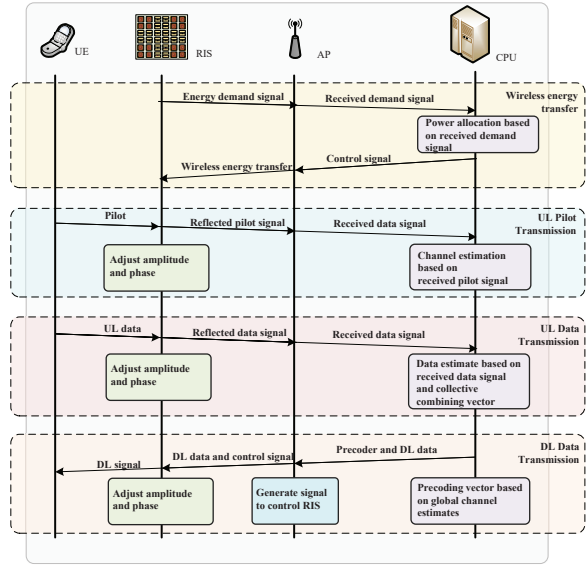}
\caption{Four-stage transmission procedure of RIS-aided CF massive MIMO systems with WET.}
\end{figure}

\subsection{System Architecture}

As shown in Fig. 2, the system is based on the CF architecture with an ``additional" RIS layer between the user equipments (UEs) and the APs. The existing CF massive MIMO system is a three-tier structure composed of the CPU, APs, and UEs. Our proposed system architecture realizes a $3.5$-layer architecture, which adds a cascading link through RIS. \textcolor{blue}{When there is a direct path, the APs can receive the signal from the UEs via two uplink paths: the direct link and the aggregated link through the RIS. As such, we treat the channels through the RIS with an additional 0.5 layer. In particular, when the direct path is blocked, this proposed structure can still guarantee stable user communication.}

The CPU has an extremely high computational capability, which not only can receive or transmit a large amount of information but also process the large number of AP receiving data \cite{zhang2021improving}. In contrast, the AP has limited computing resources which are equipped with a simple radio frequency antenna. In the considered system, the difference between the proposed RIS  and the traditional one is that our proposed one allows RIS to be equipped with a wireless energy harvesting module, which replaces the original wired-power circuit. By adopting the WET technology, AP is considered as an energy source to charge the RIS module wirelessly to ensure the normal operation of RIS.

To realize this paradigm, we introduce the flow of the four-stage transmission procedure of the system in Fig. 3. Stage \RNum{1}: Downlink energy transmission: The CPU collects the information transmitted by the AP via the fronthaul and sends control commands to the AP after signal processing. Based on the received signal, AP decides whether to transmit wireless energy signals to the RIS for energy harvesting. Especially, when the stored energy level in the RIS exceeds a threshold value, the RIS controller transmits a feedback signal to the AP for terminating the energy transmission.
Stage \RNum{2}: Data transmission: In the uplink, the UE first sends a pilot signal to the APs which reaches the APs directly or through the RISs cascaded channel. Then the APs receive the pilot signal and convey it to the CPU via dedicated fronthaul for channel estimation. Subsequently, the UE sends their uplink data and the APs receive the signal via a direct path as well as the reflected signal arriving through RISs beamforming. Besides, the APs send their received data to the CPU, which performs joint signal processing for data detection. Finally, the CPU obtains the uplink decoded signal based on the global channel estimation. Then the downlink data signal is generated by the downlink precoder and transmitted to the AP, which finally reaches the UEs through the RIS. Meanwhile, AP generates the signal to control the RIS for phases adaptation. \textcolor{blue}{Note that if the AP desires to realize dynamic control to RIS, it is necessary to modify the frame structure and to insert some control time slots. Correspondingly, deploying signal processing modules at RIS may be needed to respond to the control signals.}

\section{Deployment Design}\label{se:performance}
In this section, we propose a RIS hardware design scheme supported by the WET technology. On this basis, we explore different system operation modes and compare their advantages and disadvantages. Finally, we discuss various practical scenarios on how to effectively determine the location of RIS deployments in practice.

\subsection{Hardware Design}
The hardware implementation of RIS is based on the concept of ``metasurface", each element of which is a programmable sub-wavelength structural unit composed of two-dimensional metamaterials \cite{liaskos2018new}. In practice, the field-programmable gate array (FPGA) can be used as a controller to achieve flexible control of the RIS, which usually communicates and coordinates with other network components (e.g., BS, APs, and UEs) through dedicated links.
Although FPGA consumes a small amount of energy, it still needs some electrical power source to support its operation.
As shown in Fig. 4, we introduce the wireless energy scavenging module to the original RIS panel and exploit some elements for energy reception and other elements for signal reflection.
\textcolor{blue}{The energy harvesting elements are connected with a piece of energy storage hardware (e.g., a rechargeable battery), which can store the harvested energy and support the energy consumption of other elements performing reflection.}

RIS is generally assembled by hundreds of elements \cite{wu2019towards}, so it is worth exploring which elements are selected for serving as energy receiving modules. Here, we have designed and compared three types of hardware structures in Fig. 4.
\textcolor{blue}{The first type consists of a complete separation of the energy harvesting elements and the information reflecting elements which are easy to implement in hardware. Yet, when the RIS panel is large, there would be some energy reception dead spots in this design due to non-uniform energy flux created by impinging signals. Based on this, we further propose an improved block structure, i.e., deployment energy harvesting elements at the four corners of the RIS, i.e., type 2, which can alleviate the impacts of dead spots but introduce moderate hardware implementation difficulty. Finally, we also design a scattered structure that aims at reducing the impacts caused by dead zones for energy harvesting or signal reflection, but at the cost of higher hardware complexity.}
\textcolor{blue}{In practice, the energy consumption of the centralized and distributed element designs mainly depends on the number of elements used for information reflection.}
\textcolor{blue}{Indeed, in addition to considering the balance between the system performance and the implementation complexity, the ratio between the number of RIS elements in energy harvesting mode and that in reflecting mode is another key issue. This is determined by various factors, such as distance, energy conversion efficiency, and the channel environment. In practice, the information and energy elements ratios can be adaptively adjusted according to the feedback of the actual parameters to realize dynamic assignment \cite{9483903}.}

\begin{figure*}[ht]
\label{Figure 4}
\centering
\includegraphics[scale=0.65]{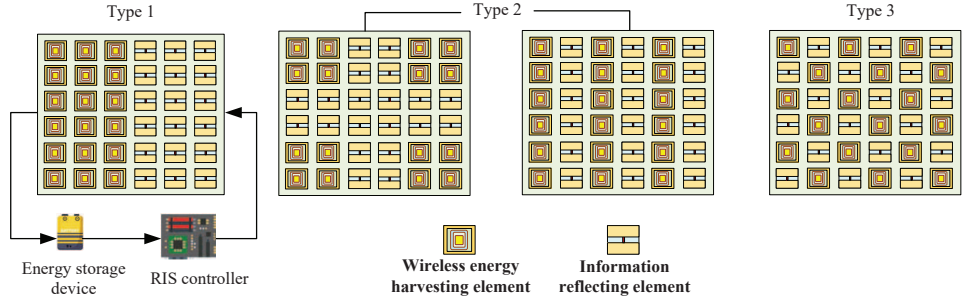}
\caption{Hardware designs for information and energy transmission in a RIS. Type 1: Energy harvesting and information reflecting elements are completely separated and distributed in blocks. Type 2: Block structure with incomplete separation of energy harvesting and information reflection elements. Type 3: Energy harvesting and information reflecting elements alternate with each other in a scattered distribution.  }
\end{figure*}

\subsection{Operation Modes}
In practice, there are different ways for deploying RISs leading to different system operation modes.
They affect the hardware design complexity and signal processing methods, as well as the synergy between individual communication devices, which inevitably brings different system performances.
In this section, we discuss several different operation modes of RIS for WET and information reflection.
\subsubsection{Centralized RIS}
As shown in Fig. 6, we first consider that there is only a single large-scale centralized RIS with massive elements. It is configured with high computational power and signal processing capability at the RIS controller, which can reduce the processing burden of the AP. Meanwhile, the centralized RIS has high beamforming capability that can assist the APs to service massive UEs via excellent interference management \cite{9483903}. Moreover, the centralized design facilitates highly efficient wireless energy focusing, while it is easier to deploy and maintain the equipment.
However, due to the architecture of centralized RIS, the path loss of the long-distance devices can be severe as UEs are randomly deployed. Also, the ability of WET to keep such a sophisticated controller functioning properly still needs to be explored. Besides, the requirements for hardware design are also demanding resulting in high cost.

\subsubsection{Non-cooperative Distributed RIS}
In this case, we assume that multiple RISs have non-cooperatively deployed in a service area. This distributed deployment is more flexible and can avoid dead spots in WET as much as possible.
Meanwhile, the distance between RIS-AP and RIS-UE is shortened that reducing the path loss of WET. As such, significant improvements in energy efficiency and communication quality can be achieved.
In practice, the hardware design of each RIS is still relatively simple since it only expects to implement simple signal reflection.
Besides, since multiple non-cooperative RISs are distributed in the service area, the signals between different RISs may interfere with each other, which degrades the communication performance of UEs.

\subsubsection{Cooperative Distributed RIS}
Similar to the last operation mode, we consider that there are multiple RISs in a service area. The difference is that there is some intelligent cooperation among RISs through physical links or dedicated wireless communication as shown in Fig. 6.
\textcolor{blue}{Note that when there is a physical link connected to the RIS, WET is regarded as supplementary electric energy to reduce the required energy consumption.}
This operation mode not only reduces the path loss of WET and information communication but also enables better beam management among multiple RISs through cooperation thus reducing interference.
However, it also places additional requirements on the hardware design of RISs, such as how to implement physical connections between multiple RISs and the design of the controller.
\textcolor{blue}{As such, RIS will generate additional energy consumption for channel estimation and cooperation. However, due to the stable characteristics of the channel, frequent communication between RISs is not required, so the required energy consumption is still reasonable.}
In practice, a high computational requirement at the CPU and advanced optimization algorithms are needed to achieve efficient and intelligent cooperation of RISs.
\textcolor{blue}{In practice, centralized RIS is more suitable for small-scale networks with dense users. In contrast, distributed RIS has better performance in large-scale networks due to its inherent ability in exploiting spatial diversities.}

\textcolor{blue}{In Fig. 5, we compare different RIS operation modes for different element numbers within an area of 1000 $\text{m}^2$ and consider the effect of phase errors. The achievable rate (bps/Hz) and the transmit power (dBm) are the two main performance metrics for information communication. It shows that RIS can significantly improve the system performance and compared with centralized RIS, distributed deployment can obtain higher system achievable rate. Moreover, phase errors reduce the performance of the system and the influence of phase errors is more significant with the increase of the RIS element numbers.}

\begin{figure}[t]
\centering
\includegraphics[scale=0.45]{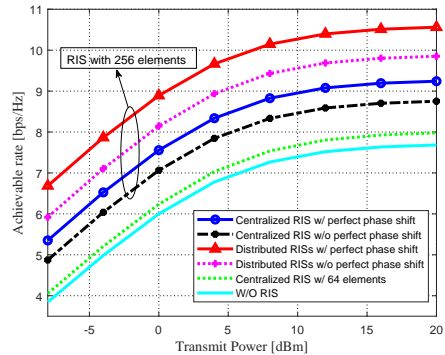}
\caption{Achievable rate of different RIS operation modes and elements with phase errors.
\label{Figure_5}}
\end{figure}

\begin{figure*}[ht]
\label{Figure 6}
\centering
\includegraphics[scale=0.8]{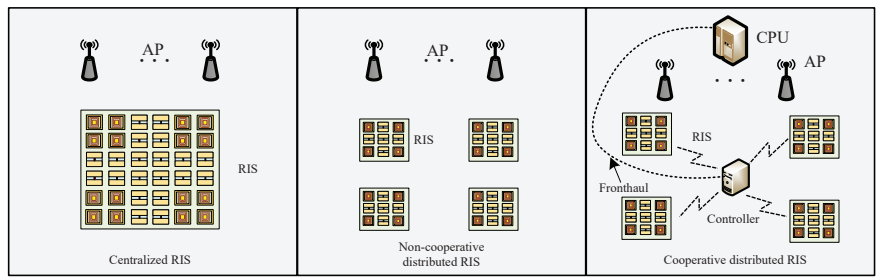}
\caption{Comparison of different RIS operation modes.}
\end{figure*}

\subsection{RIS Deployment}
How to judiciously deploy multiple RISs in a hybrid wireless network containing passive RISs, active CPUs, and APs to optimize the system performance is a critical problem that needs to be solved.
Previous studies have provided some solutions for RIS location deployment, such as deploying a RIS close to the UE or the AP side can obtain higher system performance \cite{liu2020ris}. Unfortunately, the considered system is not a single point-to-point communication and the RIS needs support from the WET technology. In the considered network, energy and information transmission are two equally important performance metrics.

First, from the perspective of optimizing the system communication performance, the RIS should be intuitively deployed in a location with a clear LoS from the AP to maximize its received signal power for passive beamforming and WET.
Although this deployment scheme enjoys good performance for the case of a single UE, it does not work well when the number of UEs increases, due to the lack of enough non-LoS paths to achieve high-rank MIMO channels. To enhance the experience of UEs with poor communication quality, we should also deploy some RISs in dedicated areas with weak signal coverage and communication dead zones caused by blockages.
On the other hand, from the perspective of WET, RIS should be deployed in scenarios where it is difficult to install powerlines for emergencies, such as UAV communication and temporary placement sites. In practice, ensuring proper system operation first requires that the RIS can effectively receive wireless energy signals to support its operation.
As such, a RIS should be reasonably deployed near an AP to facilitate the establishment of a LoS path.

In practice, the dynamic of propagation environment and user mobility lead to channel variations over time and each RIS may also be associated with multiple APs and UEs while multiple RISs may be interconnected.
In such cases, heuristic solutions to the design of RISs deployment may be ineffective.
In particular, it is generally intractable to acquire accurate globally channel state information (CSI) at a low cost in large-size systems. Therefore, how to determine a precise location deployment scheme of multiple RISs based on partial system information is a new problem with high practical significance.
A promising approach to solve this problem is to utilize machine learning techniques, such as deep learning (DL).
For example, in the training phase, we can empirically deploy multiple RISs at reasonable reference locations and collect key performance metrics, such as the received signal strength and the corresponding energy efficiency measured at different UE locations. The collected key metrics and the location of the RIS are then exploited to train the DL-based neural network as an input and an output, respectively.
\textcolor{blue}{However, during the training process, it is generally not possible to change the locations of RISs and additional installation costs will be incurred. In fact, the obtained results serve as a performance upper bound for a fixed RIS deployment location. As for more practical implementation, one may consider statical long-term optimization.}
\textcolor{blue}{Besides, we can adopt convex optimization methods such as gradient descent method and Newton method to optimize RIS location and system power allocation to improve the system performance.}

\section{Future Directions}\label{se:numerical}
In addition to the system architecture and deployment schemes discussed above, there still are some other research directions worthy of future investigation for the systems.

\subsection{Control Scheme}
\textcolor{blue}{For a single RIS, how to control the activation of energy harvesting and information reflection elements is an important problem to be solved with the following aspects. For instance, stringent requirements are imposed for the hardware design of RIS while higher information receiving and processing capabilities are also required. Besides, the massive elements in RIS incur new challenges to the optimization algorithm of high-dimensional matrices \cite{9483903}.}
As for the systems, the traditional analysis results in the literature do not consider the cooperation among RISs, which may lead to UE unfairness and communication quality degradation due to signal interference among multiple RISs.
\textcolor{blue}{To fully utilize the advantage of the strong directivity of the RIS reflected beam, the RIS phase shift matrix is designed according to the AP and UE positions to achieve precise control of the beam direction, thereby avoiding interference as much as possible.}
\textcolor{blue}{For example, with a large number of APs and RISs, computationally-efficient distributed machine learning algorithms serve as appealing solutions as they have been successfully applied to address large-scale optimization problems.}
However, a rigorous control protocol must be designed to ensure orderly communication and harness the interference among devices.

\subsection{Resource Allocation}
Large and complex systems inevitably face the problem of resource allocation.
In traditional CF systems, pilot assignment is an important research direction that is also inherent to our system. Besides, for the consideration of supporting WET technology, power allocation is another key issue worthy of attention.
Specifically, we need to control the energy supply offered by APs to optimize the system energy efficiency.
If all the APs adopt a uniform power allocation to supply energy to RISs, it is generally suboptimal which can cause energy waste.
As such, certain optimization mechanisms are needed to maximize power utilization.
\textcolor{blue}{For instance, for imperfect CSI, we can exploit the large-scale channel information and solve the related resource allocation problem through optimization algorithms such as geometric programming.}
In practice, formulating design optimization problems to realize accurate and efficient power allocation is an ideal solution. However, most of the practical problems are non-convex and intractable in complex networks.
Hence, finding an optimal power control scheme within a reasonable timescale is an urgent future research direction.
\textcolor{blue}{It is worth mentioning that using the WET technology to supply energy to UEs is also worth considering in resource allocation.}

\subsection{Hardware}
To achieve ubiquitous deployment of RIS-aided CF massive MIMO systems, the cost and quality of hardware for receivers and transmitters as well as RIS naturally swing a non-trivial trade-off in system design.
Yet, to overcome the associated high cost, we typically utilize low-cost components at APs.
\textcolor{blue}{Unfortunately, the use of low-cost hardware at RIS affects the hardware accuracy, including the limited phase shift resolution and the mutual coupling of phase and incidence angles in each RIS element. In practice, the non-linear amplitude and phase response of RIS would affect the accuracy of reflected signals, resulting in unsatisfactory system performance. As a result, a pragmatic system design taking into the potential hardware imperfect is necessary.} Besides, the hardware design of energy reception introduces extra hardware complexity in fabricating RIS. Indeed, designing WET elements at the wavelength level is challenging. In addition, how to avoid the coupling interference between the WET elements and the reflecting elements is also a problem that needs to be considered.
On the other hand,  the fronthaul capacity limitation reduces the availability of converging control signals between the APs and the CPU.
As such, the above hardware impairment problems would definitely degrade the communication and energy transmission performance of the system that are important issues for investigation.

\section{Conclusions}\label{se:conclusion}
In this article, we investigate the promising RIS-aided CF massive MIMO system with WET technology for realizing IoE in future wireless networks.
First, we discussed several potential application scenarios of the system and the proposed system architecture.
Besides, we proposed different operation modes and shed light on the suitable RIS hardware design for WET.
In addition, we investigated feasible solutions for the possible deployment of RISs in the system.
Finally, to offer useful guidance for future research, we indicated the critical challenges and promising research directions for realizing RIS-aided CF massive MIMO systems with WET.

\bibliographystyle{IEEEtran}
\bibliography{IEEEabrv,Ref}

\end{document}